# ARTICLE

# Formation of HNC and HCN isomers in molecular plasmas revealed by frequency comb and quantum cascade laser spectroscopy

Ibrahim Sadiek,*[a] Simona Di Bernardo,[a,c,d] Uwe Macherius[b] and Jean-Pierre H. van Helden [a]

Hydrogen cyanide (HCN) is a well-known product in combustion, astrophysical, and plasma environments, but its isomer, hydrogen isocyanide (HNC), remains unexplored in molecular plasmas. Here, we report on the detection and quantification of both HNC and HCN isomers in low-temperature plasmas generated from $N_2/H_2/CH_4$ mixtures using quantum cascade laser and frequency comb absorption spectroscopy. The identification of HNC is confirmed by comparison with molecular spectroscopic databases. The observed [HNC]/[HCN] abundance ratio of $\sim 10^{-4}$ is far lower than reported values in interstellar media, where ratios can reach unity. We attribute this stark difference to fundamentally different kinetics, dominated by an interplay between formation of vibrationally 'hot' intermediates (HCN* and HNC*) and their subsequent stabilization and destruction. This mechanism contrasts with cold interstellar environments, where dissociative recombination reactions dominate and the formed HNC persists and accumulates over astrochemical timescales. Our results highlight the overlooked role of HNC in plasma chemistry containing carbon, nitrogen and hydrogen, with implications for plasma-assisted HCN synthesis and nitrogen-carbon plasma processes in surface modification and material treatments.

## Introduction

Plasma synthesis of hydrogen cyanide (HCN) directly from methane ($CH_4$) and nitrogen ($N_2$) has been suggested quite recently as a method for on demand and catalyst-free production at ambient pressure.[1,2] As an alternative to conventional thermal methods,[3,4] this approach leverages the highly reactive nature of non-equilibrium plasmas, which enables molecular activation at low temperatures, effectively bypassing limitations of thermodynamic equilibrium.[5-7] Furthermore, HCN has been reported as a major product in several low-temperature molecular plasmas involving hydrogen ($H_2$), $N_2$, and a carbon source,[8,9] as well as in processes such as plasma-assisted nitrocarburizing.[10,11] In such environments, the chemistry of HCN is believed to be of central importance in material processing. However, its isomer, hydrogen isocyanide (HNC) — an even more reactive species — has never been conclusively detected in these plasmas, and its role remains unexplored. With a ground state energy of 14.97 kcal/mol [12] (or 16.31 kcal/mol[13] from a different study) higher than HCN and a barrier height of 43.82 kcal/mol separating the two isomers,[14] HNC - if formed - could significantly influence the yield of HCN by serving as a major exit channel in its potential plasma-assisted production; and affect plasma compositions and enhance surface reactions in material processing applications.

Beyond molecular plasma applications, HNC is a key intermediate in several astrochemical and combustion environments.[15-22] In cold interstellar media (ISM), observed [HNC]/[HCN] abundance ratios range from 0.01 to values above unity[17,18] — far higher than thermochemical equilibrium predictions, which suggest a ratio of $\sim 1.9 \times 10^{-33}$ at 100 K (using the recent energy difference value of 14.97 kcal/mol[12]). This discrepancy led to the suggestion that dissociative recombination (DR) of protonated hydrogen cyanide, $HCNH^+$ with cold electrons plays a decisive role in the mechanism of isomeric fractionation,[15] which was later supported by phase-space and quantum chemical calculations,[23,24] and experimentally validated by Mendes et al.[25] using event-by-event fragment momentum spectroscopy. Their results revealed that the DR of $HCNH^+ + e^-$, which proceeds with an energy excess of 136.1 kcal/mol above HCN + H,[14] yields HCN and HNC in nearly equal amounts.

In combustion systems, HCN is the dominant cyanide species and a key contributor to $NO_x$ formation during the combustion of fossil fuels.[20-22] Although less abundant, HNC is suggested to be an important intermediate in the oxidation of HCN. At elevated temperatures, the two isomers rapidly equilibrate, and HNC, due to its higher reactivity, is thought to provide a faster route to cyanide oxidation.[26]

Early spectroscopic efforts to provide reference spectra for HNC were motivated primarily by astrophysical interests. The first spectroscopic measurement of HNC was made under non-equilibrium conditions using matrix isolation infrared spectroscopy at 14 K.[27] Later, Maki and Sams[28] were able to measure its gas phase infrared emission spectrum by shifting the HCN ⇌ HNC equilibrium towards HNC at ~1000 K. Picqué et

[a.] *Experimental Physics V: Spectroscopy of Atoms and Molecules by Laser Methods, Faculty of Physics and Astronomy, Ruhr University Bochum, 44780 Bochum, Germany, E-mail: ibrahim.sadiek@ruhr-uni-bochum.de*
[b.] *Leibniz Institute for Plasma Science and Technology (INP), 17489 Greifswald, Germany*
[c.] *Department of Mathematics and Physics, University of Campania "Luigi Vanvitelli", 81100 Caserta, Italy*
[d.] *Italian Aerospace Research Centre (CIRA), 81043 Capua, Italy.*
† Electronic supplementary information (ESI) available. See DOI: 10.1039/x0xx00000x

al.[29] measured the emission spectra of HCN/HNC in a radio-frequency plasmas containing $H_2$ and $N_2$. Despite using only $H_2$ and $N_2$ as precursors, both HCN and a very week signal of HNC were detected, attributed to carbon impurities in their reactor. Subsequent laboratory work on reference samples led to the acquisition of extensive emission spectra for HCN[30-33] and HNC.[34-36] These efforts led to the identification of over 40,000 rotation-vibration energy levels for both isomers so far.[13,37-39]

Here, we apply optical frequency comb and quantum cascade laser absorption spectroscopy to investigate the {H, C, N} system chemistry in plasma processes containing $N_2$, $H_2$, and $CH_4$ as precursors. This dual-diagnostic approach enables excitation of the strong fundamental rovibrational bands: the $\nu_3$ band of HNC near 2080 cm$^{-1}$, and the $\nu_1$ band of HCN near 3300 cm$^{-1}$. As a result, the detection of trace amounts of reactive HNC becomes possible when it forms under plasma conditions using multipass optics. From the measured absorption spectra, we accurately determined the [HNC]/[HCN] abundance ratio. A time series measurements revealed a steady fourfold increase in HCN within the first ~40 minutes after plasma ignition. The asynchronous fluctuations of the two isomers during the preconditioning of the plasma chamber indicate that catalytic isomerization reactions, converting HNC to HCN, play a dominant role. We developed a kinetic scheme and analytical model incorporating key formation, isomerization, and destruction pathways under plasma conditions. The model includes two possible formation routes—hot and cold channels—and demonstrates the strong kinetic suppression of HNC in molecular plasmas. It also shows that the [HNC]/[HCN] abundance ratio is highly sensitive to parameters such as internal energy relaxation and destruction rates.

## Experimental

Fig. 1 illustrates the overall experimental setup, which consists of (i) the plasma reactor, (ii) an optical frequency comb-based Fourier transform spectrometer (OFC-FTS), and (iii) an external-cavity quantum cascade laser (EC-QCL) absorption spectrometer.

The plasma was generated in a DC discharge, operated at a pressure of (1.00 ± 0.01) mbar, using gas flows of 11 standard cubic centimeter (sccm) $N_2$, 1 sccm $H_2$, and 8 sccm $CH_4$, and a plasma power of 700 W. The reactor was a custom-built, scaled-down version of an industrial plasma reactor used for nitrocarburizing stainless-steel and tool materials. All measurements were performed in the plasma chamber after preconditioning (with the plasma being on) for 1.5 hours. Plasma nitrocarburizing is used here as a model system for {H,N,C}-containing plasmas, wherein HCN and its more reactive isomer, HNC, are anticipated to play a dominant role in plasma chemistry.

The frequency comb setup used a mid-infrared (mid-IR) frequency comb (Menlo Systems) as a light source in combination with a home-built FTS. The mid-IR output of ~ 150 mW, spanning from 2800 – 3400 cm$^{-1}$, was produced through non-linear difference frequency generation process in a periodically poled lithium niobate crystal. The comb had a repetition frequency, $f_{rep}$, of 250 MHz, which was locked to the output of a tunable direct digital synthesizer referenced to a GPS-referenced oscillator. Details of the reactor and the comb system are available elsewhere.[40,41]

High-resolution spectra were obtained using the sub-nominal data analysis approach,[42,43] which has been applied previously to measure complex spectra of halogenated molecules,[44-46] istopopocules of $N_2O$,[47] and for plasma diagnostics.[11] In this method, the nominal resolution of the FTS is matched to the $f_{rep}$ of the comb, which ensures precise sampling of the intensities of the comb modes, and enables absolute frequency calibration. Measurements were performed at four different $f_{rep}$ values, each offset by 62 MHz in the optical domain, to accurately sample the individual absorption profiles. Full details of the OFC-FTS setup can be found elsewhere.[11]

The EC-QCL system covered the 1985 – 2250 cm$^{-1}$ range and included two liquid nitrogen cooled photo detectors – PD1 and PD2 for the signal and reference arms, respectively. Wavenumber calibration was performed using reference gas cells of CO (1 mbar) and $CO_2$ (50 mbar), and a Ge etalon. Details of the spectral acquisition and frequency calibration of EC-QCL measured spectra are included in Fig. S1 and Fig. S2 in the ESI.†

## Results and discussion

### Comb spectroscopy of $N_2/H_2/CH_4$ molecular plasmas

The OFC-FTS setup captured the entire comb spectrum in a single acquisition, which enabled us to identify prominent C-H and N-H absorbing species formed in the plasma in the 2800 – 3400 cm$^{-1}$ range. As shown in Fig. 2a, highly resolved rovibrational features could be measured for $CH_4$, $C_2H_2$, HCN, and $NH_3$. For the HCN isomer, we could measure a total of five vibrational bands. Fig. 2b shows a zoomed-in spectral window of the P($J$ = 24) absorption transition in the fundamental $\nu_1 \leftarrow \nu_0$ band of HCN, where $J$ is the rotational quantum number, along with a Gaussian line shape fit. A flat residual is shown in the lower panel, indicating the high precision of the absorption profile measurements. The fitted profiles were used to determine the populations of rovibrational states and to construct Boltzmann plots, as shown Fig. 2c, from which rotational temperatures were determined.

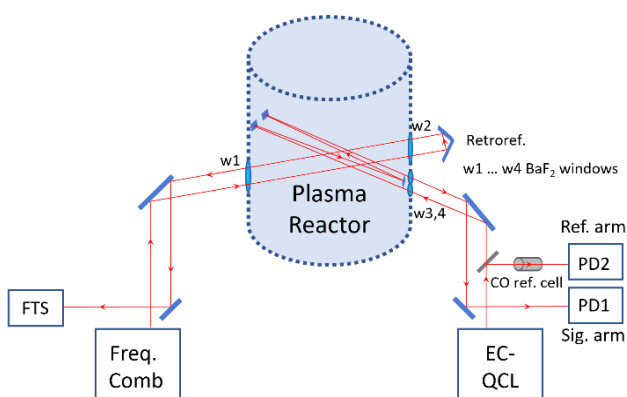

**Fig. 1** Overall experimental setup, consisting of (i) the plasma reactor, (ii) a comb-based Fourier transform spectrometer (FTS), and (iii) an EC-QCL-based spectrometer with liquid nitrogen cooled photodetectors, PD1 and PD2 for the signal and reference arms, respectively. The comb beam passed twice through the reactor, resulting in an effective absorption pathlength of (146.0 ± 0.5) cm, while the EC-QCL beam was coupled into a multi-pass optics of White cell arrangements, resulting in an effective absorption pathlength of (1310.4 ± 3.2) cm. The optical windows were made of wedged $BaF_2$ materials.

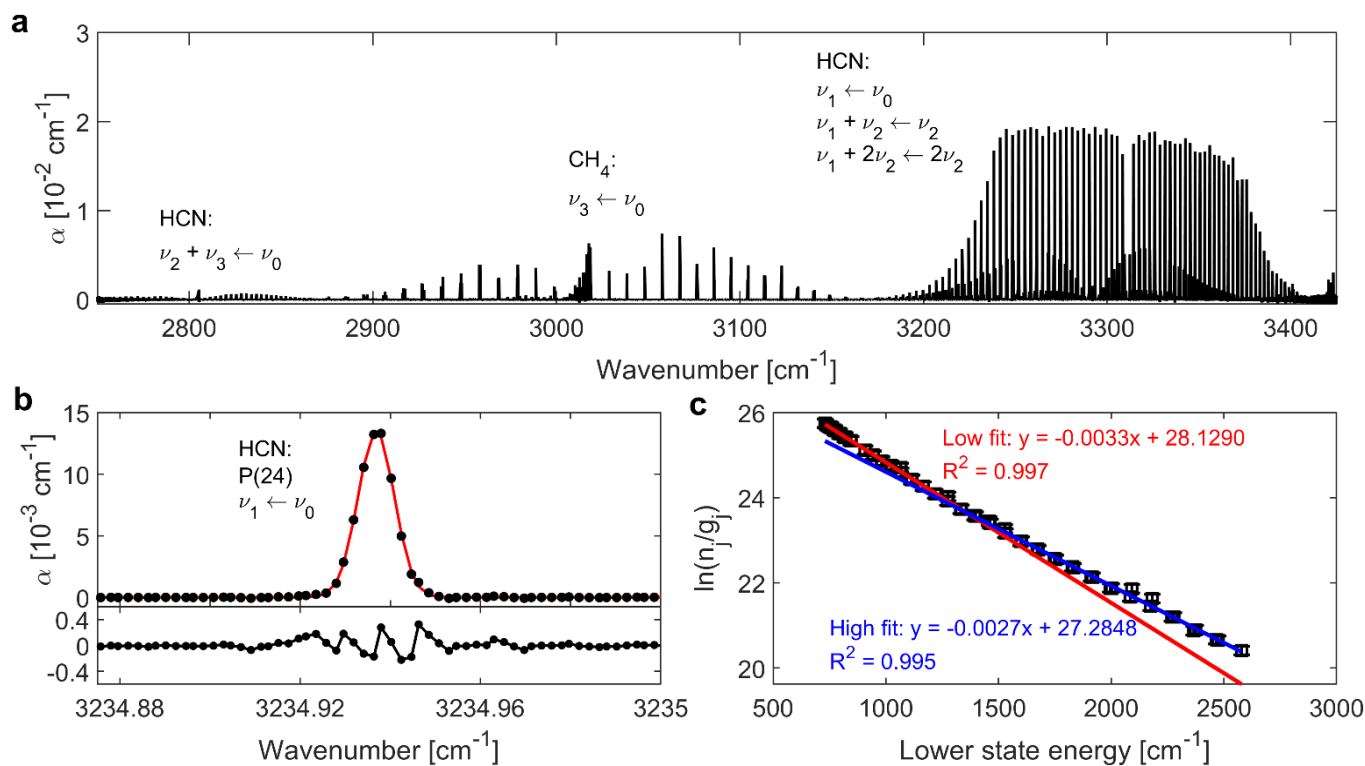

**Fig. 2** Measured high-resolution absorption coefficient, a, of molecular species generated in plasmas with 11 sccm $N_2$, 1 sccm $H_2$, and 8 sccm $CH_4$ gas flows at a total pressure of (1.00 ± 0.01) mbar, and a plasma power of 700 W, (a); Absorption profile of the P(24) transition in the $n_1 \leftarrow n_0$ band of $H^{12}C^{14}N$, together with a fit using a Gaussian line shape function, (b); and Non-local thermodynamic equilibrium (non-LTE) of populations of rovibrational levels indicated by the two different linear slopes for levels with lower (red linear fit) and higher (blue linear fit) lower state energy than ~1400 $cm^{-1}$, (c). The two slopes resulted in two distinct temperatures of 436 K and 533 K with a difference of ~97 K for the $11^10 \leftarrow 01^10$ vibrational hot band.

Our measurements revealed clear non-local thermodynamic equilibrium (non-LTE): Two distinct rotational temperatures of 436 K and 533 K were found, differing by ~97 K, for transitions with lower state energies above and below ~1400 $cm^{-1}$ (corresponding to a crossing point, $J_{cross}$ of 21) in the $11^10 \leftarrow 01^10$ vibrational hot band. Similar behavior for other vibrational bands was observed but at a different value of $J_{cross}$. From these rotational population measurements, we determined the total populations of the $00^00$, $01^10$, and $02^00$ vibrational states. The details of the population determination from the experimental spectrum are provided in ESI.†

**Identification of reactive HNC intermediate**

To identify whether HNC forms and to quantify it, we used the EC-QCL setup to probe its strong $v_3$ rovibrational transitions. Fig. 3 shows representative absorption spectrum near 2081.5 $cm^{-1}$ measured by the EC-QCL setup. The spectrum features clearly resolved transitions from HCN and CO, as confirmed by comparison with the HITRAN2020 database.[48] One additional feature was identified as the R(20) transition of the $00^01 \leftarrow 00^00$ band of HNC (see the inset of Fig. 3). There is no reference line list for the HNC isomer in the HITRAN database,[48] however, there is an improved line list for HCN/HNC molecular system[13] in the ExoMol database.[49] Our measured transition at 2082.077 $cm^{-1}$ is in very good agreement with the R(20) transition of HNC in the ExoMol database within $4 \times 10^{-4}$ $cm^{-1}$ in the spectral position. In total, we have measured eight rovibrational transitions in the 2069 – 2088 $cm^{-1}$ range which we attribute to the HNC isomer. The measured line centers showed excellent agreement with the updated ExoMol line list of HNC,[13] and with reference infrared Fourier transform spectroscopy (FTIR) measurements on thermally heated sample of HCN (to shift the equilibrium to HNC).[50] These results are summarized in Table 1, which also includes the signal-to-noise ratio (SNR), the fitting uncertainty for each transition frequency, and the difference in transition frequencies between measurements and ExoMol line lists.[49] The full set of measured R-branch absorption profiles ($J$ = 15 – 22) is shown in Fig. S3 in the ESI.†

**Table 1** Transition frequencies assigned to HNC and their corresponding values from EXoMol database,[49] and FTIR reference measurements.[50] The number in parentheses represent a 1s uncertainty from the fit using a Gaussian line shape function. The SNR of each measured profile is also listed. The difference in transition frequencies between our EC-QCL measurements and the ExoMol line list is given in the final column (a-b) x $10^{-4}$.

| | Transition frequency [$cm^{-1}$] | | | SNR | (a – b) |
|---|---|---|---|---|---|
| | This work[(a)] | ExoMol[(b),49] | FTIR[50] | | × $10^4$ |
| R(15) | 2069.1472(4) | 2069.1506(4) | 2069.1506 | 8 | -34 |
| R(16) | 2071.7790(3) | 2071.7837(4) | 2071.7840 | 12 | -47 |
| R(17) | 2074.3900(2) | 2074.3930(4) | 2074.3930 | 14 | -30 |
| R(18) | 2076.9815(3) | 2076.9784(4) | 2076.9784 | 12 | 31 |
| R(19) | 2079.5444(5) | 2079.5400(4) | 2079.5402 | 6 | 44 |
| R(20) | 2082.0780(5) | 2082.0776(4) | -- | 7 | 4 |
| R(21) | 2084.5913(3) | 2084.5910(4) | 2084.5909 | 11 | 3 |
| R(22) | 2087.0809(4) | 2087.0804(4) | 2087.0806 | 10 | 5 |

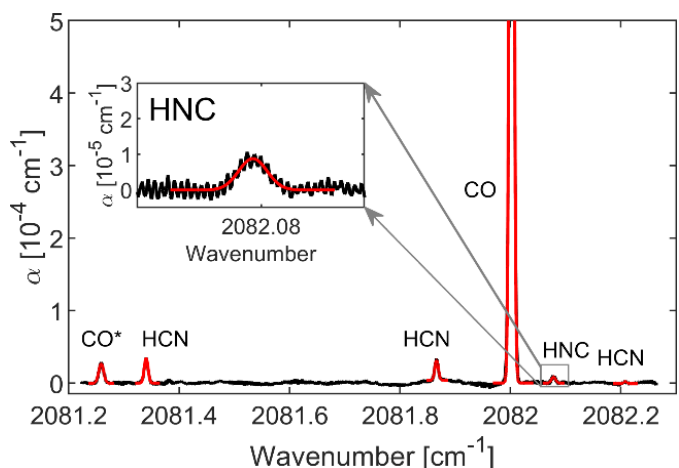

**Fig. 3** Measured absorption coefficient, α, of plasma-generated species using the EC-QCL setup, together with the Gaussian fits (red) to the resolved transitions. The measured spectrum contains absorption profiles for HCN and CO (and overpopulated hot band transition of CO near 2081.26 cm$^{-1}$, denoted as CO*). The identified transition of HNC, around 2082.077 cm$^{-1}$, is shown in the inset together with a Gaussian line shape fit.

The present measurements provide unambiguous spectroscopic evidence for the formation of HNC under the investigated plasma conditions. Owing to its substantially higher chemical reactivity relative to HCN, the [HNC]/[HCN] abundance ratio represents a sensitive diagnostic of plasma reactivity and a potential indicator for tailoring plasma operating conditions to specific application requirements, as discussed below.

## [HNC]/[HCN] abundance ratio in molecular plasmas

To determine the [HNC]/[HCN] abundance ratio from spectroscopic measurements, we follow four approaches. (*i*) We determined the abundance ratio from populations of rotational states. We used the lower-state population of the R(20) rovibrational transition of HNC at 2082.077 cm$^{-1}$, measured using an EC-QCL, and compared it to the corresponding transition of HCN obtained from broadband OFC-FTS. This approach takes advantage of the broad spectral coverage of the comb, allowing population comparisons between rotational states with identical *J* quantum numbers, ensuring accuracy in determined abundance ratio, particularly under non-LTE conditions of molecular plasmas. (*ii*) We determined the ratio from the total population of the ground vibrational states. For HNC, we first determined the rotational populations from measured profiles for transitions listed in Table 1, determined the rotational temperature from a Boltzmann plot, and then extrapolated to determine the total population of the 00$^0$0 vibrational state. For HCN, whose rovibrational populations show non-LTE (see Fig. 2c), we employed a bimodal temperature model to estimate the total population in its ground vibrational state. (*iii*) We determined the ratio by assuming that HNC follows the same bimodal temperature distribution as HCN, justified by the similarity in their formation mechanisms — both largely produced through exothermic radical-radical reactions that impart significant internal excitation (*see below*) before relaxing to their steady-state concentrations. Finally, (*iv*) for comparative purposes, we estimated the ratio from densities determined by assuming thermal equilibrium at the gas kinetic temperature, derived from the Doppler widths of the measured profiles. Detailed population analysis is provided in the ESI.†

**Table 2** Populations (in molecule cm$^{-3}$) of HNC and HCN from EC-QCL and OFC-FTS measurements, respectively, together with the corresponding [HNC]/[HCN] abundance ratio. The uncertainties are evaluated by propagating the errors originating from fit uncertainties.

| | (*i*) | (*ii*) | (*iii*) | (*iv*) |
|---|---|---|---|---|
| HNC × 10$^{-11}$ | 0.11 ± 0.02 | 5.23 ± 1.98 | 2.07 ± 0.17 | 4.80 ± 0.64 |
| HCN × 10$^{-15}$ | 0.05 ± 0.01 | 4.55 ± 0.09 | 4.55 ± 0.09 | 3.08 ± 0.11 |
| [HNC]/[HCN] × 10$^4$ | 2.15 ± 0.32 | 1.15 ± 0.80 | 2.20 ± 0.10 | 1.56 ± 0.30 |

(*i*): lower-state populations of the R(20) lines for both HNC and HCN
(*ii*): ground vibrational state populations. For HNC based on the determined rotational temperature of transitions of Table 1. For HCN, a bi-modal $T_{rot}$ approach is used to evaluate the non-LTE populations.
(*iii*): ground vibrational state populations assuming a bi-modal $T_{rot}$ for both HNC and HCN.
(*iv*): assuming thermal population at the gas kinetic temperatures

Table 2 lists the evaluated populations based on the different scenarios (*i*) – (*iv*) and the resulting ratios. Overall, the [HNC]/[HCN] ratio in molecular plasmas lies in the range of (1.15– 2.20) × 10$^{-4}$. As the EC-QCL measurements (see Fig. 3) included some transitions of HCN (the P(5) transition of the 00$^0$1 ← 00$^0$0 band at 2081.863 cm$^{-1}$, and the P(11) transition of the 03$^1$0 ← 00$^0$0 band at 2082.206 cm$^{-1}$), we evaluated the population of HCN based on those transitions as well. The agreement between EC-QCL and OFC-FTS data, including a cross-comparison of shared transitions, lies within 25 − 30%, which is within the combined uncertainties of the two measurement methods.

## Time series measurements

A time series experiment using the EC-QCL setup was performed to monitor the temporal evolution of HNC and HCN steady-state populations from plasma ignition up to 120 minutes (during the preconditioning phase). Fig. 4 presents the measured densities of HNC and HCN (left axis, logarithmic scale), along with the [HNC]/[HCN] abundance ratio (green, right axis). At plasma onset (0.5 min), the HNC density is initially high (~1.5×10$^{12}$ cm$^{-3}$), while HCN starts near ~1×10$^{15}$ cm$^{-3}$. Over the first 10 minutes, HNC decreases by approximately a factor of 1.5 and then exhibits mild fluctuations, reflecting possible plasma instabilities or drifts in operational parameters. In contrast, HCN accumulates steadily, increasing nearly fourfold to reach a steady-state value of ~4.6×10$^{15}$ cm$^{-3}$ within about 40 minutes. Actually, the undulations observed in the HNC density are also present in the HCN time profile but out of phase. These asynchronous oscillations also echoed in the [HNC]/[HCN] abundance ratio curve (green, Fig. 4), suggesting that the two isomers respond to plasma dynamics through distinct yet interlinked chemical pathways.

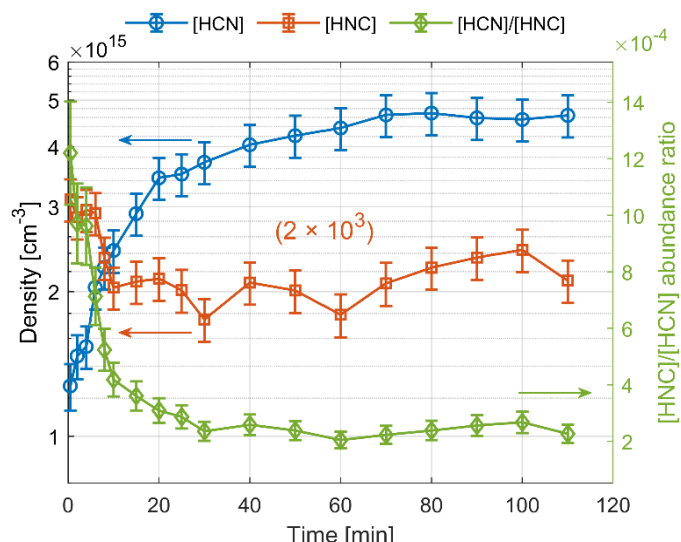

**Fig. 4** Time series measurements of HCN (blue circles) and HNC (red squares) densities using the EC-QCL setup – left axis, together with [HNC]/[HCN] abundance ratio (green diamonds) – right axis.

Rather than simple parallel formation or destruction, the opposing behavior points to an overall dynamic interconversion mechanism, where changes in one species are mirrored by changes in the other. This is consistent with rapid isomerization reactions occurring on collision-limited timescales, likely catalyzed by reactive species, such as atomic hydrogen, carbon, or oxygen (H, C, or O). Such catalytic processes can efficiently convert HNC to HCN in response to minor perturbations in plasma conditions, including shifts in electron temperature, power, or radical concentrations. It is worth noting that the dual diagnostic setup developed here enables quantification of densities of the two isomers over a range spanning four orders of magnitude, demonstrating the very wide dynamic range of the setup.

The [HNC]/[HCN] abundance ratio begins at ~12×10$^{-4}$ and drops within the first 10–20 minutes, stabilizing near 2 ×10$^{-4}$ after ~40 minutes. Importantly, this decline is primarily driven by the accumulation of HCN, most likely by the very fast catalytic reactions of HNC + M → HCN + M with M being another species such as H, C, or O adsorbed on the surface or in the plasma. Further time-resolved measurements on microsecond to millisecond timescales would be valuable for capturing transient kinetic responses. Such measurements will be capable of capturing the predicted fast decrease in HNC abundance after plasma ignition, and it will also capture its instantaneous responses to changes in concentrations of reactive species such as CH, C, and $H_3^+$.

**Kinetic scheme and analytical model**

The measured [HNC]/[HCN] abundance ratio in molecular plasmas is significantly lower than values typically reported in cold molecular clouds and star-forming regions,[15,17,18] where the ratio can approach or even exceed unity, despite HCN being the more thermodynamically stable isomer. In ISM, this high ratio is commonly attributed to the DR of HCNH$^+$ with cold electrons (HCNH$^+$ + e$^-$ → HCN + H or HNC + H), a pathway that efficiently populates both isomers,[14] and leads to a near-statistical distribution between HCN and HNC.[25] This is justified as the DR reaction exhibits a high cross-section at low electron energies typical for cold ISM. However, other radical-radical and ion-radical reactions also contribute to isomer formation in the ISM.[51] Even at slightly higher temperatures in hot cores and young stellar objects,[52,53] the formation starts to favor HCN over HNC due to the onset of HNC destruction via reactions with atomic H, C, and O, albeit at slow rates compared to plasma environments. Amano et al.[54] further demonstrated this trend in an emission spectroscopy study using an extended negative glow discharge at 77 K (higher than typical cold molecular clouds of temperatures of 10 – 50 K), where the measured [HNC]/[HCN] abundance ratio already deviated significantly from ISM conditions. This indicates that alternative formation mechanisms, next to the DR, become operative even at such relatively low temperatures.

In industrial molecular plasmas, however, HCN – and, as demonstrated here for the first time, also HNC – are produced from various precursor gas mixtures (e.g., $N_2/H_2/CH_4$), but the resulting [HNC]/[HCN] ratio is markedly lower in the order of 10$^{-4}$. This stark difference suggests a fundamentally different set of chemical pathways. In plasmas, fast electron-driven processes such as ionization and excitation occur at rates far exceeding those of DR. The DR cross-section for HCNH$^+$ declines sharply with increasing electron temperature —from about 2 × 10$^{-11}$ cm$^2$ at 0.01 meV to ~3 × 10$^{-15}$ at 100 meV[14] — and it becomes even negligible at the electron temperatures 1 – 10 eV typical for molecular plasmas. At these energies, the HCNH$^+$ intermediate itself becomes short-lived and susceptible to fragmentation or charge exchange with neutrals like $H_2$ and $N_2$, neither of which yields HCN or HNC. This diminishes the effective HCNH$^+$ concentration and further shifts the chemistry away from DR-mediated pathways.

Instead, exothermic radical-radical reactions are predicted to supersede the formation of HCN and HNC in plasmas. These 'hot' formation pathways generate internally excited molecules (HCN* and HNC*) carrying substantial internal energy stored in the form of vibrational and rotational excitation. For example, the reactions: (N + $CH_2$ → HCN + H, $\Delta H$ = −133.7 kcal/mol) and $NH_2$ + C → HNC + H, $\Delta H$ = −118.5 kcal/mol) are highly exothermic. Herbst et al.[55] suggested that the largest fraction of this exothermicity will be transferred into internal energy, estimated at 103.9 kcal/mol for HNC and 113.6 kcal/mol for HCN, predominantly stored in vibrational and rotational modes, and the minor rest will be translated as motion. This is consistent with our comb spectroscopy measurements of the gas kinetic temperature (~1 kcal/mol or 500 K), obtained from Doppler broadening of the absorption profiles. Similarly, for the reaction NH + CH → HNC + H or HCN + H, an even higher exothermicity of 143.2 kcal/mol and 130.5.5 kcal/mol was calculated for the HCN and HNC pathways, respectively.[51]

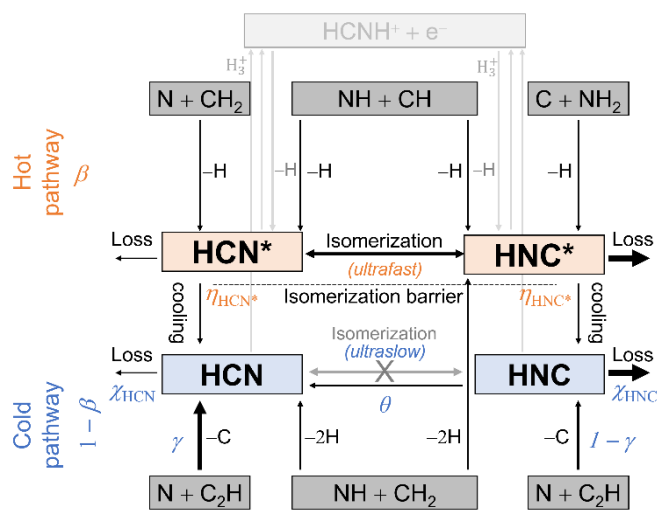

**Fig. 5** Schematic of major chemical processes involved in the determination of the final [HNC]/[HCN] abundance ratio in molecular plasmas.

In Fig. 5, we present a schematic of the key reaction pathways and processes governing the [HNC]/[HCN] abundance ratio in molecular plasmas. Central to this scheme is the behavior of the internally excited intermediates, HNC$^*$ and HCN$^*$, which can undergo ultrafast unimolecular isomerization prior to vibrational relaxation. When the internal energy of these intermediates exceeds the isomerization barrier (~43.8 kcal/mol for HCN → HNC and ~ 28.9 kcal/mol HNC → HCN), the interconversion rate can reach ~ $10^{13}$ s$^{-1}$. Such fast isomerization will overwrite any memory of the initial formation pathways and effectively compensate for losses in any of the internally exited isomers, if it happens, before relaxation. Similar behavior has been reported for isomerization involving light hydrogen moiety, such as in H$_2$NC$^+$ ⇌ HCNH$^+$.[56] Thus, the fast isomerization between HCN$^*$ and HNC$^*$ is expected to establish a near-instantaneous equilibrium on the sub-picosecond timescale, long before significant vibrational cooling occurs. However, as the molecular system approaches the isomerization barrier, the rate slows to ~ $10^{11}$ s$^{-1}$.[56] A critical population of molecules exists with an internal energy approximately equal to the transition state barrier; in this regime, the instantaneous [HCN$^*$]/[HNC$^*$] ratio reflects the isomerization rate near the barrier energy. Once the internal energies drop below the isomerization threshold, other processes such as collisional (or radiative, in the case of ISM) relaxation and dissociation dominate.

In addition to these hot channels, this schematic also includes a pathway where the formation directly yields 'cold' products with an internal energy below the isomerization barriers. This can be the case for both isomers or for only one of them. For example, radical-radical reactions N + C$_2$H → HCN + C, (∆H = −44.7 kcal/mol) and N + C$_2$H → HNC + C, (∆H = −32.0 kcal/mol).[57] Although the exothermicity of these reactions still appears larger than the isomerization barriers (~43.8 kcal/mol starting from the HCN side and ~ 28.9 kcal/mol starting from the

HNC side), the small portion that is transferred to translational motion will cause the internal energy of both HCN$^*$ and HNC$^*$ to fall just below the barrier. In this case, relaxation or dissociation processes will compete with isomerization until the molecules become frozen in the ground state with almost negligible direct unimolecular isomerization. However, 'catalytic' isomerization of reactive HNC in the ground state (or partially excited) can still convert it to HCN. For example, the reaction HNC + C → HCN + C is barrierless or has only a small activation barrier (~few hundred Kelvin), proceeding at a near collision-limited rate (1.15 × 10$^{-10}$ cm$^3$ molecule$^{-1}$ s$^{-1}$),[57] as estimated from capture rate theory. Similarly, the reaction CH + HNC is exothermic and takes place at a rate of 1.40 ×10$^{-10}$ cm³/molecule$^{-1}$ s$^{-1}$.[58] As the abundance of C and CH in such plasma nitrocarburizing plasmas is predicted to be high, particularly at high plasma powers, these reactions will effectively convert HNC into HCN, increasing its budget.

Another illustrative case are the NH + CH$_2$ → HNC$^*$ +2H (∆H = −133.8 kcal/mol) and NH + CH$_2$ → HCN +2H (∆H = −41.8 kcal/mol) reactions. The former will form hot HNC$^*$ first, while the latter will directly form cold HCN. This reaction underscores the formation of isomers with different internal energy profiles, which further support the observed low [HNC]/[HCN] abundance ratio in our experiments. In Table S2 in ESI,† we list exothermic chemical reactions that act as a potential formation pathway for both hot and cold HCN and HNC isomers.

Building on the kinetic framework illustrated in Fig. 5, we developed an analytical model that semi-quantitatively describes the formation and fate of HCN and HNC isomers in plasmas, and it explains the observed lower [HNC]/[HCN] abundance ratio in terms of dominant molecular and kinetic processes. The aim of this exercise was to provide chemical kinetic parameters that could ideally be linked to experimental variables for tuning the HNC abundance based on the application needs for high reactive HNC (relevant to material processing applications like plasma nitrocarburizing) or efficient HCN production (potential plasma-assisted chemical synthesis of HCN). The model incorporates both 'hot' and 'cold' formation channels, as well as subsequent isomerization, relaxation, and destruction processes. In this model, a fraction of the total production, $\beta$ proceeds via hot formation pathways. This fraction rapidly interconverts through isomerization and are assumed to establish a transient equilibrium before relaxation. Assuming symmetry in isomerization, we approximate [HCN$^*$] ≈ [HNC$^*$] ≈ $\beta$/2. The remaining fraction (1−$\beta$) corresponds to cold formation, where internal energy is insufficient to allow efficient unimolecular isomerization. It also accounts for other channels that bypass the excited state pathway entirely and form cold HCN via endothermic reactions, e.g., NH$_3$ + C → HCN + H$_2$, which is a key step in conventional thermal synthesis of HCN and requires 1100 K.[59] One can expect that most of these cold contributions go to HCN, as formation of cold HNC is less favorable due to its higher energy and reactivity. Nevertheless, the cold channel is further portioned into $\gamma$ for cold HCN and (1

− γ) for cold HNC formation. Excited HCN* and HNC*, once dropped below the isomerization barrier, can either relax to ground-state molecules or be lost through destruction before being fully relaxed. The probability of successful relaxation to HCN and HNC is captured by the cooling efficiencies $\eta_{HCN^*}$ and $\eta_{HNC^*}$, respectively. Additionally, ground-state HNC is subject to further transformation: it may be destroyed (efficiency $\chi_{HNC}$) or converted to HCN (fraction $\theta$), reflecting its higher chemical reactivity even in the ground state. HCN, on the other hand, is generally more stable and tends to persist under plasma conditions, see Fig. 4, therefore it is better to define the corresponding term $\chi_{HCN}$ as stabilization or accumulation efficiency (see ESI†).

Assuming total production, $P_{tot} = 1$, the final steady state abundances of [HCN] and [HNC] are given by:

$$[HCN] = (1 - \chi_{HCN}) \cdot \left[\frac{\beta}{2} \cdot \eta_{HCN^*} + \gamma \cdot (1 - \beta)\right] + \theta \cdot \left[\frac{\beta}{2} \cdot \eta_{HNC^*} + (1 - \gamma) \cdot (1 - \beta)\right] \quad (1)$$

$$[HNC] = (1 - \chi_{HNC} - \theta) \cdot \left[\frac{\beta}{2} \cdot \eta_{HNC^*} + (1 - \gamma) \cdot (1 - \beta)\right] \quad (2)$$

The complete derivation of the model and the final expression for the steady-state abundance ratio are given in ESI,† together with expression for the ratio in the special cases of hot-pathway-dominated chemistry (relevant to non-thermal molecular plasmas) and cold-pathway-dominated chemistry (relevant to conventional thermal processes).

This model reveals that the steady-state [HNC]/[HCN] abundance ratio is governed by a finely balanced interplay between formation of excited intermediates and their relaxation, as well as the conversion of HNC to HCN and the destruction processes. The relaxation efficiency of excited HNC* ($\eta_{HNC^*}$) plays a crucial role in enabling HNC survival, especially under high $\beta$ conditions where hot formation dominates akin to plasma conditions. Conversely, the destruction efficiency ($\chi_{HNC}$) and the conversion fraction to HCN ($\theta$) act strongly to suppress HNC, and their combined influence becomes critical when ($\chi_{HNC} + \theta$) ≈ 1, a regime associated with extreme HNC depletion as observed in plasma experiments. Cold formation preferentially favors HCN due to energetic constraints, with a marginal (1 − γ) contribution to cold HNC formation. Parameter sensitivity surface plots (see Fig. S5 in see the ESI†) demonstrate sharp gradients in the [HNC]/[HCN] abundance ratio, with steep cliffs and valleys in response to small variations in key parameters, highlighting the sensitivity of HNC abundance in reactive environments.

The [HNC]/[HCN] abundance ratio can serve as a valuable indicator for optimizing process yields in plasma-based applications. In plasma nitrocarburizing, a higher ratio—favoring the more reactive HNC isomer—may enhance the overall plasma reactivity, potentially improving the diffusion of reactive species into treated surfaces. Conversely, in potential plasma-assisted HCN synthesis, a lower [HNC]/[HCN] ratio may be more desirable, as HNC acts as a competing destruction channel that can reduce HCN yield. In both scenarios, the abundance ratio offers a pathway for process control. By linking experimental variables—such as gas composition, plasma power, and process pressure—to the model parameters (e.g., $\beta$, $\eta_{HNC^*}$, $\chi_{HNC}$, and $\theta$), it becomes possible to tune operating conditions with greater precision. This approach moves beyond conventional trial-and-error methods for processing control, enabling more informed and efficient plasma processes optimization.

## Conclusions

We report on the first identification and quantification of HNC in molecular plasmas containing $N_2/H_2/CH_4$, highlighting its role as a highly reactive and transient species in such environments. HNC is likely a major reactive sink that can affect the overall yield of HCN in plasma-assisted synthesis schemes.[2] Even trace carbon impurities can facilitate the formation of both HCN and HNC in $N_2/H_2$ plasmas.[29] Using a combination of broadband optical frequency comb-based Fourier-transform spectroscopy and high-resolution EC-QCL absorption spectroscopy, we performed detailed spectroscopic analysis of both isomers. By comparing spectral features to database references and constructing Boltzmann plots under non-LTE conditions, we accurately determined the [HNC]/[HCN] abundance ratio. The consistently low ratio of (1.15–2.20) × $10^{-4}$ stands in stark contrast to the ISM, where the two isomers often approach equal abundance, despite HCN being thermodynamically favored. The time series measurements of HNC and HCN reveal a steady fourfold accumulation of HCN within the first ~40 minutes after plasma ignition. In contrast, the sharp initial decrease in HNC (not fully captured here), along with its asynchronous fluctuations relative to HCN in response to plasma instabilities, points to a dominant role of catalytic isomerization reactions that interconvert HNC to HCN. We developed a kinetic scheme and analytical model that accounts for key formation, isomerization, and destruction pathways in plasma conditions. Our model considers the two possible formation pathways including the hot and cold channels. In contrast to the ISM, the radical-radical reactions dominate over dissociative recombination under molecular plasma conditions in the hot formation pathway, leading to highly vibrationally excited intermediates (HCN* and HNC*) that rapidly isomerize prior to vibrational cooling. The model highlights the strong kinetic suppression of HNC in plasmas and emphasizes how sensitive the [HNC]/[HCN] abundance ratio is to parameters such as internal energy relaxation and destruction processes. These parameters could be linked to experimental variables from gas composition to plasma power and process pressure. For example, there should be an optimal $N_2/CH_4$ ratio that yields maximum HNC yield, which could be beneficial for plasma surface processing applications, or the suppression of undesirable products for plasma-assisted HCN synthesis. Similarly, high plasma power increases the electron temperature and enhances further dissociation of $CH_4$ and $N_2$, boosting CN and CH radical formation which may selectively enhance the formation of one isomer over the other as well. Overall, we propose the isomeric ratio as a powerful diagnostic of [H, C, N] non-equilibrium plasma chemistry, with direct implications for both potential plasma-assisted synthesis of HCN, and current industrial practices such as plasma nitrocarburizing or the deposition of diamond-like films.

## Author contributions

Ibrahim Sadiek: conceptualization, methodology, software, validation, formal analysis, investigation, writing – original draft, review & editing, visualization, funding acquisition. Simona Di Bernardo: formal analysis, validation, review & editing. Uwe Macherius: investigation, methodology, formal analysis, review & editing. Jean-Pierre H. van Helden: conceptualization, methodology, review & editing, validation, resources.

## Conflicts of interest

There are no conflicts to declare.

## Data availability

All the additional experimental, spectroscopic and theoretical data are available in ESI.†

## Acknowledgements

This research was supported by the German Research Foundations (DFG-Deutsche Forschungsgemeinschaft) – project No. 499280974.

## References

1   L. Henderson, P. Shukla, V. Rudolph and G. Duckworth, *Ind. Eng. Chem. Res.*, 2020, **59**, 21347-21358.
2   N. S. Kamarinopoulou, G. R. Wittreich and D. G. Vlachos, *Sci. Adv.*, 2024, **10**, No. eadl4246.
3   E. Gail, R. Kulzer, J. Lorösch, A. Rubo, M. Sauer, R. Kellens, J. Reddy, N. Steier and W. Hasenpusch, Cyano compounds, inorganic, in Ullmann's Encyclopedia of Industrial Chemistry, 7th edn, vol. 40, Wiley-VCH, Weinheim, 2011.
4   B. E. Blackwell, C. K. Fallon, G. S. Kirby, M. Mehdizadeh, T. A. Koch, C. J. Pereira and S. K. Sengupta, Induction-heated reactors for gas-phase catalyzed reactions, US Patent US7070743B2, 2006.
5   A. Fridman, *Plasma Chemistry*, Cambridge University Press, Cambridge, 2008.
6   J. Meichsner, M. Schmidt, R. Schneider and H.-E. Wagner, Nonthermal Plasma Chemistry and Physics, CRC Press, Boca Raton, FL, 2013.
7   A. Bogaerts and E. C. Neyts, *ACS Energy Lett.*, 2018, **3**, 1013-1027.
8   C. J. Tang, I. Abe, A. J. S. Fernandes, M. A. Neto, L. P. Gu, S. Pereira, H. Ye, X. F. Jiang and J. L. Pinto, *Diam. Relat. Mater.*, 2011, **20**, 304-309.
9   D. Dekkar, A. Puth, E. Bisceglia, P. W. P. Moreira, A. V. Pipa, G. Lombardi, J. Röpcke, J. H. van Helden and F. Bénédic, *J. Phys. D: Appl. Phys.*, 2020, **53**, 455204.
10  A. Puth, L. Kusýn, A. V. Pipa, I. Burlacov, A. Dalke, S. Hamann, J. H. van Helden, H. Biermann and J. Röpcke, *Plasma Sources Sci. Technol.*, 2020, **29**, 035001.
11  I. Sadiek, A. Puth, G. Kowzan, A. Nishiyama, S.-J. Klose, J. Röpcke, N. Lang, P. Masłowski and J. H. van Helden, *Plasma Sources Sci. Technol.*, 2024, **33**, 075011.
12  T. L. Nguyen, J. H. Baraban, B. Ruscic and J. F. Stanton, *J. Phys. Chem. A*, 2015, **119**, 10929-10934.
13  R. J. Barber, J. K. Strange, C. Hill, O. L. Polyansky, G. C. Mellau, S. N. Yurchenko and J. Tennyson, *Mon. Not. R. Astron.*, 2013, **437**, 1828-1835.
14  J. Semaniak, B. F. Minaev, A. M. Derkatch, F. Hellberg, A. Neau, S. Rosen, R. Thomas, M. Larsson, H. Danared, A. Paal and M. AF Ugglas, *Astrophys. J., Suppl. Ser.*, 2001, **135**, 275-283.
15  W. D. Watson, *Rev. Mod. Phys.*, 1976, **48**, 513-552.
16  G. L. Blackman, R. D. Brown, P. D. Godfrey and H. I. Gunn, *Nature*, 1976, **261**, 395-396.
17  W. M. Irvine and F. P. Schloerb, *Astrophys. J.*, 1984, **282**, 516.
18  T. Hirota, S. Yamamoto, H. Mikami and M. Ohishi, *Astrophys. J.*, 1998, **503**, 717-728.
19  G. J. Harris, Y. V. Pavlenko, H. R. A. Jones and J. Tennyson, *Mon. Not. R. Astron. Soc.*, 2003, **344**, 1107-1118.
20  P. Dagaut, P. Glarborg and M. U. Alzueta, *Prog. Energy Combust. Sci.*, 2008, **34**, 1-46.
21  N. Lamoureux, P. Desgroux, M. Olzmann and G. Friedrichs, *Prog. Energy Combust. Sci.*, 2021, **87**, 100940.
22  M. Stuhr and G. Friedrichs, *J. Phys. Chem. A*, 2022, **126**, 9485-9496.
23  E. Herbst, *Astrophys. J.*, 1978, **222**, 508.
24  D. Talbi and Y. Ellinger, *Chem. Phys. Lett.*, 1998, **288**, 155-164.
25  M. B. Mendes, H. Buhr, M. H. Berg, M. Froese, M. Grieser, O. Heber, B. Jordon-Thaden, C. Krantz, O. Novotný, S. Novotny, D. A. Orlov, A. Petrignani, M. L. Rappaport, R. Repnow, D. Schwalm, A. Shornikov, J. Stützel, D. Zajfman and A. Wolf, *Astrophys. J.*, 2012, **746**, L8.
26  P. Glarborg and P. Marshall, *Energy Fuels*, 2017, **31**, 2156-2163.
27  D. E. Milligan and M. E. Jacox, *J. Chem. Phys.*, 1963, **39**, 712-715.
28  A. G. Maki and R. L. Sams, *J. Chem. Phys.*, 1981, **75**, 4178-4182.
29  N. Picqué and G. Guelachvili, *Spectrochim. Acta A Mol. Biomol. Spectrosc.*, 2000, **56**, 681-702.
30  G. C. Mellau, B. P. Winnewisser and M. Winnewisser, *J. Mol. Spectrosc.*, 2008, **249**, 23-42.
31  G. C. Mellau, *J. Chem. Phys.*, 2011, **134**, 194302.
32  G. C. Mellau, *J. Chem. Phys.*, 2011, **134**, 234303.
33  G. C. Mellau, *J. Mol. Spectrosc.*, 2011, **269**, 12-20.
34  G. C. Mellau, *J. Chem. Phys.*, 2010, **133**, 164303.
35  G. C. Mellau, *J. Mol. Spectrosc.*, 2010, **264**, 2-9.
36  G. C. Mellau, *J. Mol. Spectrosc.*, 2011, **269**, 77-85.
37  G. J. Harris, O. L. Polyansky and J. Tennyson, *Astrophys. J.*, 2002, **578**, 657.
38  G. J. Harris, J. Tennyson, B. M. Kaminsky, Y. V. Pavlenko and H. R. A. Jones, *Mon. Not. R. Astron. Soc.*, 2006, **367**, 400-406.

39  T. Delahaye, R. Armante, N. A. Scott, N. Jacquinet-Husson, A. Chédin, L. Crépeau, C. Crevoisier, V. Douet, A. Perrin, A. Barbe, V. Boudon, A. Campargue, L. H. Coudert, V. Ebert, J. M. Flaud, R. R. Gamache, D. Jacquemart, A. Jolly, F. Kwabia Tchana, A. Kyuberis, G. Li, O. M. Lyulin, L. Manceron, S. Mikhailenko, N. Moazzen-Ahmadi, H. S. P. Müller, O. V. Naumenko, A. Nikitin, V. I. Perevalov, C. Richard, E. Starikova, S. A. Tashkun, V. G. Tyuterev, J. Vander Auwera, B. Vispoel, A. Yachmenev and S. Yurchenko, *J. Mol. Spectrosc.*, 2021, **380**, 111510.
40  I. Sadiek, A. J. Fleisher, J. Hayden, X. Huang, A. Hugi, R. Engeln, N. Lang and J. H. van Helden, *Commun. Chem.*, 2024, **7**, 110.
41  I. Sadiek, N. Lang and J. H. Van Helden, *Opt. Express*, 2024, **32**, 46511-46521.
42  P. Maslowski, K. F. Lee, A. C. Johansson, A. Khodabakhsh, G. Kowzan, L. Rutkowski, A. A. Mills, C. Mohr, J. Jiang, M. E. Fermann and A. Foltynowicz, *Phys. Rev. A*, 2016, **93**, 021802.
43  L. Rutkowski, P. Masłowski, A. C. Johansson, A. Khodabakhsh and A. Foltynowicz, *J. Quant. Spectrosc. Radiat. Transf.*, 2018, **204**, 63-73.
44  I. Sadiek, A. Hjältén, F. C. Roberts, J. H. Lehman and A. Foltynowicz, *Phys. Chem. Chem. Phys.*, 2023, **25**, 8743-8754.
45  A. Hjältén, A. Foltynowicz and I. Sadiek, *J. Quant. Spectrosc. Radiat. Transf.*, 2023, **306**, 108646.
46  I. Sadiek, A. Hjältén, F. Senna Vieira, C. Lu, M. Stuhr and A. Foltynowicz, *J. Quant. Spectrosc. Radiat. Transf.*, 2020, **255**, 107263.
47  A. Hjältén, I. Sadiek and A. Foltynowicz, *J. Quant. Spectrosc. Radiat. Transf.*, 2025, **340**, 109452.
48  I. E. Gordon, L. S. Rothman, R. J. Hargreaves, R. Hashemi, E. V. Karlovets, F. M. Skinner, E. K. Conway, C. Hill, R. V. Kochanov, Y. Tan, P. Wcisło, A. A. Finenko, K. Nelson, P. F. Bernath, M. Birk, V. Boudon, A. Campargue, K. V. Chance, A. Coustenis, B. J. Drouin, J. M. Flaud, R. R. Gamache, J. T. Hodges, D. Jacquemart, E. J. Mlawer, A. V. Nikitin, V. I. Perevalov, M. Rotger, J. Tennyson, G. C. Toon, H. Tran, V. G. Tyuterev, E. M. Adkins, A. Baker, A. Barbe, E. Canè, A. G. Császár, A. Dudaryonok, O. Egorov, A. J. Fleisher, H. Fleurbaey, A. Foltynowicz, T. Furtenbacher, J. J. Harrison, J. M. Hartmann, V. M. Horneman, X. Huang, T. Karman, J. Karns, S. Kassi, I. Kleiner, V. Kofman, F. Kwabia–Tchana, N. N. Lavrentieva, T. J. Lee, D. A. Long, A. A. Lukashevskaya, O. M. Lyulin, V. Y. Makhnev, W. Matt, S. T. Massie, M. Melosso, S. N. Mikhailenko, D. Mondelain, H. S. P. Müller, O. V. Naumenko, A. Perrin, O. L. Polyansky, E. Raddaoui, P. L. Raston, Z. D. Reed, M. Rey, C. Richard, R. Tóbiás, I. Sadiek, D. W. Schwenke, E. Starikova, K. Sung, F. Tamassia, S. A. Tashkun, J. Vander Auwera, I. A. Vasilenko, A. A. Vigasin, G. L. Villanueva, B. Vispoel, G. Wagner, A. Yachmenev and S. N. Yurchenko, *J. Quant. Spectrosc. Radiat. Transf.*, 2022, **277**, 107949.
49  J. Tennyson, S. N. Yurchenko, J. Zhang, C. A. Bowesman, R. P. Brady, J. Buldyreva, K. L. Chubb, R. R. Gamache, M. N. Gorman, E. R. Guest, C. Hill, K. Kefala, A. E. Lynas-Gray, T. M. Mellor, L. K. McKemmish, G. B. Mitev, I. I. Mizus, A. Owens, Z. Peng, A. N. Perri, M. Pezzella, O. L. Polyansky, Q. Qu, M. Semenov, O. Smola, A. Solokov, W. Somogyi, A. Upadhyay, S. O. M. Wright and N. F. Zobov, *J. Quant. Spectrosc. Radiat. Transf.*, 2024, **326**, 109083.
50  J. B. Burkholder, A. Sinha, P. D. Hammer and C. J. Howard, *J. Mol. Spectrosc.*, 1987, **126**, 72-77.
51  J.-C. Loison, V. Wakelam and K. M. Hickson, *Mon. Not. R. Astron. Soc.*, 2014, **443**, 398-410.
52  P. Schilke, C. M. Walmsley, G. Pineau Des Forets, E. Roueff, D. R. Flower and S. Guilloteau, *Astron. Astrophys.*, 1992, **256**, 595-612.
53  F. L. Schöier, J. K. Jørgensen, E. F. van Dishoeck and G. A. Blake, *Astron. Astrophys.*, 2002, **390**, 1001-1021.
54  T. Amano, Z. Zelinger, T. Hirao, J. Takano and R. Toyoda, *J. Mol. Spectrosc.*, 2008, **251**, 252-255.
55  E. Herbst, R. Terzieva and D. Talbi, *Mon. Not. R. Astron. Soc.*, 2000, **311**, 869-876.
56  D. Talbi and E. Herbst, *Astron. Astrophys.*, 1998, **333**, 1007-1015.
57  J. C. Loison and K. M. Hickson, *Chem. Phys. Lett.*, 2015, **635**, 174-179.
58  E. Hébrard, M. Dobrijevic, J. C. Loison, A. Bergeat and K. M. Hickson, *Astron. Astrophys.*, 2012, **541**, A21.
59  R. M. Badger, *J. Am. Chem. Soc.*, 1924, **46**, 2166-2172.